\documentclass{aa}  

\usepackage{graphicx}
\usepackage{amsmath}
\usepackage{amssymb}
\usepackage{txfonts}
\usepackage{threeparttable}
\usepackage[final]{microtype}
\usepackage{url}
\usepackage[normalem]{ulem}
\usepackage{fancybox}

\usepackage{xspace}

\def\hessj{\mbox{HESS\,J1731$-$347}\xspace}
\def \cco {CCO in HESS\,J1731$-$347\xspace}
\def \xmmu {\mbox{XMMU\,J1732}\xspace}
\def \paper {Paper\,I\xspace}

\begin{document} 

\title{The neutron star in HESS\,J1731$-$347: CCOs as laboratories to
  study the equation of state of superdense matter}

   \author{D.~Klochkov\inst{1} \and
          V.~Suleimanov\inst{1,2}\and
          G.~P\"uhlhofer\inst{1}\and
          D.~G.~Yakovlev\inst{3}
          A.~Santangelo\inst{1}\and
          K.~Werner\inst{1}
          }

   \institute{Institut f\"ur Astronomie und Astrophysik, Universit\"at
     T\"ubingen (IAAT), Sand 1, 72076 T\"ubingen, Germany
     \and
     Kazan (Volga region) Federal University,
     Kremlevskaya 18, 420008 Kazan, Russia 
     \and
     Ioffe Institute, Politekhnicheskaya 26, 194021,
     St. Petersburg, Russia 
   }

   \date{Received ***, 2014; accepted ***, 2014}

 
  \abstract
   {Central compact objects (CCOs) in supernova remnants are isolated
     thermally emitting neutron stars (NSs). They are most probably
     characterized by a
     magnetic field strength that is roughly two orders of magnitude
     lower than that of most of the radio and accreting pulsars. The
     thermal emission of CCOs can be modeled to obtain
     constraints on the physical parameters of the star such as its
     mass, radius, effective temperature, and chemical composition.}
   {The \cco is one of the brightest objects in this class. Starting
     from 2007, it was observed
     several times with different X-ray satellites. Here we present our
     analysis of two new \emph{XMM-Newton} observations of the source
     performed in 2013 which increase the total exposure time of the
     data available for spectral analysis by a factor of about five
     compared to the analyses presented before.} 
   {We use our numerical spectral models for carbon and
       hydrogen atmospheres to fit the 
     spectrum of the CCO. From our fits, we derive
     constraints on the physical parameters of the emitting star such
     as its mass, radius, distance, and effective temperature. We also
     use the new data to derive new upper limits on the source
     pulsations and to confirm the absence of a long-term flux and
     spectral variability.} 
   {The analysis shows that atmosphere models are
     clearly preferred by the fit over the blackbody spectral function. Under
     the assumption that the X-ray emission is uniformly produced by
     the entire star surface (supported by the lack of pulsations),
     hydrogen atmosphere models lead to uncomfortably large distances of
     the CCO,
     above 7--8\,kpc. On the other hand, the carbon atmosphere model
     formally excludes distances above 5--6\,kpc and is compatible with
     the source located in the Scutum-Crux ($\sim$3\,kpc) or
     Norma-Cygnus ($\sim$4.5\,kpc) Galactic spiral arm.
     We provide and discuss the corresponding confidence contours in the
     NS mass--radius plane.
     The measured effective temperature indicates that the NS is
     exceptionally hot for the estimated age of $\sim$30\,kyr. We
     discuss possible cooling scenarios to explain this property, as
     well as
     possible additional constraints on the star mass and radius from
     cooling theory.}
   {}

   \titlerunning{Neutron star in HESS~J1731$-$347 / G353.6$-$0.7}
   \maketitle

%
\section{Introduction}

More than a third of all young neutron stars (NSs) associated with
supernova remnants (SNRs) are observed as \emph{central compact
objects} or CCOs. These point-like X-ray sources are characterized by
thermal X-ray spectra with $kT\sim 0.2-0.5$\,keV and have so far not been found
to emit in other electromagnetic wavebands. 
The CCOs were first considered
as a separate class of isolated NSs after their characterization with the 
\emph{Chandra} observatory \citep{Pavlov:etal:02,Pavlov:etal:04}.
Around a dozen CCOs are known so far including
candidates. Pulsations with periods between $\sim$0.1\,s and
$\sim$0.4\,s have been detected in three of them
\citep[e.g.,][]{Gotthelf:etal:13}. Lack of magnetic activity such as 
non-thermal (magnetospheric) emission and lack of associated pulsar wind
nebula, together with the low measured spin-down rates 
or the corresponding upper limits,
lead to relatively small estimated
magnetic field strengths of the CCOs, $B\lesssim 10^{10}- 10^{11}$\,G
\citep{Halpern:Gotthelf:10,Gotthelf:etal:13}. This is roughly two orders of
magnitude below the field strengths observed in most of the radio and
accreting pulsars and in the small group of nearby X-ray emitting Dim
Isolated Neutron Stars (XDINS). Thus, CCOs
apparently represent a class of young
($\lesssim$10$^3-10^4$\,yr), low-magnetized, thermally emitting
cooling NSs.

The absence of magnetospheric and accretion phenomena
potentially provides an undisturbed view of the stellar surface of CCOs. The
geometrical and physical properties of the star can be probed through
a modeling of the spectral and timing properties of its X-ray emission.
The mass and radius of the compact star which can be estimated from
the modeling are directly related to the nature of the superdense
matter in its interior. Different theoretical models predict different
equations of state (EOS) of superdense nuclear matter and, thus,
different  mass-radius relations for NSs which can be
compared with observations \citep[e.g.,][]{Lattimer:Prakash:07}.

The CCO \object{XMMU\,J173203.3$-$344518} (to be referred as \xmmu in
the following) was discovered with \emph{XMM-Newton} in 2007 as a
point-like X-ray source roughly at the center of the TeV-emitting
supernova remnant 
\object{HESS~J1731$-$347}, also known as \object{G\,353.6$-$0.7}
\citep{Acero:etal:09,Tian:etal:10,HESS:2011}.  The \emph{XMM-Newton}
observations and the subsequent observations
with \emph{Suzaku} \citep{Bamba:etal:12} and \textsl{Chandra} have
shown that the source exhibits a blackbody-like X-ray 
spectrum with absorption at low energies characterized by $kT\simeq 0.5$\,keV 
(typical for a CCO)
and an absorption column density $n_{\rm H}\simeq 1.5\times 10^{22}$\,cm$^{-2}$ 
\citep{Acero:etal:09,Halpern:Gotthelf:10:b,Bamba:etal:12}.

No X-ray pulsations have so far been found in \xmmu. Our recent analysis
of the \emph{XMM-Newton} timing observations in 2012 presented in 
\citet{Klochkov:etal:13}, hereafter \paper, yielded an upper
limit of $\sim$10\% on the pulsed fraction of sinusoidal pulsations.
Although the limit is not very strong, it can indicate that
the X-ray emission is roughly uniformly produced by
the entire stellar surface. This assumption was adopted in \paper to
couple the emitting area with that of the compact star.  \citet{HESS:2011}
argued that the remnant is most likely located either in the
Scutum-Crux arm or in the 
Norma-Cygnus arm, with the corresponding spiral arm distances of
$\sim$3\,kpc or $\sim$4.5\,kpc following the model by
\citet{Hou:etal:09}. They also derived a lower limit on the distance
of $\sim$3.2\,kpc based on the measured X-ray absorption and the $^{12}$CO
emission in the direction of the remnant. The fit of the CCO
spectrum with an absorbed blackbody model under an assumption of a
canonical NS radius of 10\,km leads to an unrealistic
distance estimate of $\sim$30\,kpc. 
A solution to a similar problem in case of the non-pulsating CCO
in Cas~A, whose distance is well known,
was proposed by \citet{Ho:Heinke:09} \citep[see, however,][for an
alternative explanation]{Pavlov:Luna:09}. The usage of a carbon atmosphere
spectral model allowed \citet{Ho:Heinke:09} to
reconcile the distance to the source with the radius of a canonical
NS.
Following a similar approach, the carbon
atmosphere models described in \citet{Suleimanov:etal:14} 
have been applied to the \cco and allowed 
the estimated distance to the remnant of $\sim$3$-$4\,kpc to be
reconciled with the 
canonical radius of the neutron star as described in \paper. 

Here we present the analysis of two new observations of \xmmu
performed with \emph{XMM-Newton} in 2013 (PI: G.~P\"uhlhofer) which
increase the total exposure time of the observations
suitable for an accurate spectral analysis by a factor of about five.
The new data put much stronger constraints on the source spectral
continuum. We apply our carbon and hydrogen atmosphere models to the
new spectra and derive new constraints on the physical parameters of
the emitting star such as its mass and radius. 
We discuss possible cooling scenarios of the NS based on the
measured effective temperature and the available age estimate. The
cooling theory under certain assumptions allows us to put additional
constraints on the star mass and radius.
The paper is organized as follows. The available observations and
the long term monitoring of the source flux are described in
Sect.~\ref{sec:data}. A new upper limit on the source pulsations is
provided in Sect.~\ref{sec:uplim}. Spectral analysis is described
in  Sect.~\ref{sec:spe}. The spectral fits using the carbon and  
hydrogen atmosphere models are presented in
Sects.\,\ref{sec:carbon} and \ref{sec:hydrogen}, respectively.
The cooling history of the NS and the corresponding constraints on the
mass and radius of the NS are discussed in Sect.\,\ref{sec:cooling}.

\section{Observational data and the long-term flux history of
  XMMU\,J173203.3$-$344518\label{sec:data}}

Since 2007, the \cco has regularly been observed with the
\emph{XMM-Newton, Chandra, Suzaku,} and \emph{Swift} satellites. Most
of the observations are targeted at the supernova 
remnant \hessj. A short log of the available observations is
provided in  Table\,\ref{tab:obs}. In this work, we focus on the
analysis of the last two observations 
performed in 2013
with the \emph{XMM-Newton} orbital observatory \citep{Jansen:etal:01}. We
refer to them as the March observation and the October 
observation in the following. To combine the data with those taken in
the 2007 \emph{XMM-Newton} observations, we also re-processed the 2007 observations
using the latest pipeline version and calibration (see next Section).

\begin{table}
  \begin{centering}
  \caption{Summary of observations of 
    XMMU\,J173203.3$-$344518.}
  \label{tab:obs}
  \begin{tabular}{l l l l}
    \hline\hline
    Date & Satellite/ & Exposure & Time res.\\
         &  obs. mode &  [ksec]  &          \\
    \hline
                &                   &    &     \\
    2007 Feb 23$^{(1)}$ & \emph{Suzaku}/imaging   & 41 & 8\,s\\
    2007 Mar 21$^{(2)}$ & \emph{XMM}/imaging & 25 & 70\,ms \\
    2008 Apr 28 & \emph{Chandra}/imaging    & 30 & 3.2\,s\\
    2009 Feb 4 & \emph{Swift}/imaging       & 1.4& 2.5\,s\\
    2009 Mar 9 & \emph{Swift}/imaging       & 1.4& 2.5\,s\\
    2010 May 18$^{(3)}$  & \emph{Chandra}/timing    & 40 & 2.85\,ms\\
    2012 Mar 2$^{(2)}$ & \emph{XMM}/timing  & 24 & 0.03\,ms \\
    2013 Mar 7$^{(*)}$ & \emph{XMM}/imaging  & 72 & 70\,ms \\
    2013 Oct 6$^{(*)}$ & \emph{XMM}/imaging  & 61 & 70\,ms \\
    \hline
  \end{tabular} 
  \end{centering}
  $^{(1)}$\citet{Bamba:etal:12}\\
  $^{(2)}$\citet{Klochkov:etal:13}\\
  $^{(3)}$\citet{Halpern:Gotthelf:10c}\\
  $^{(*)}$The new observations analyzed in this work.
  \footnotesize 
\end{table}

\begin{figure}
\centering
\resizebox{\hsize}{!}{\includegraphics{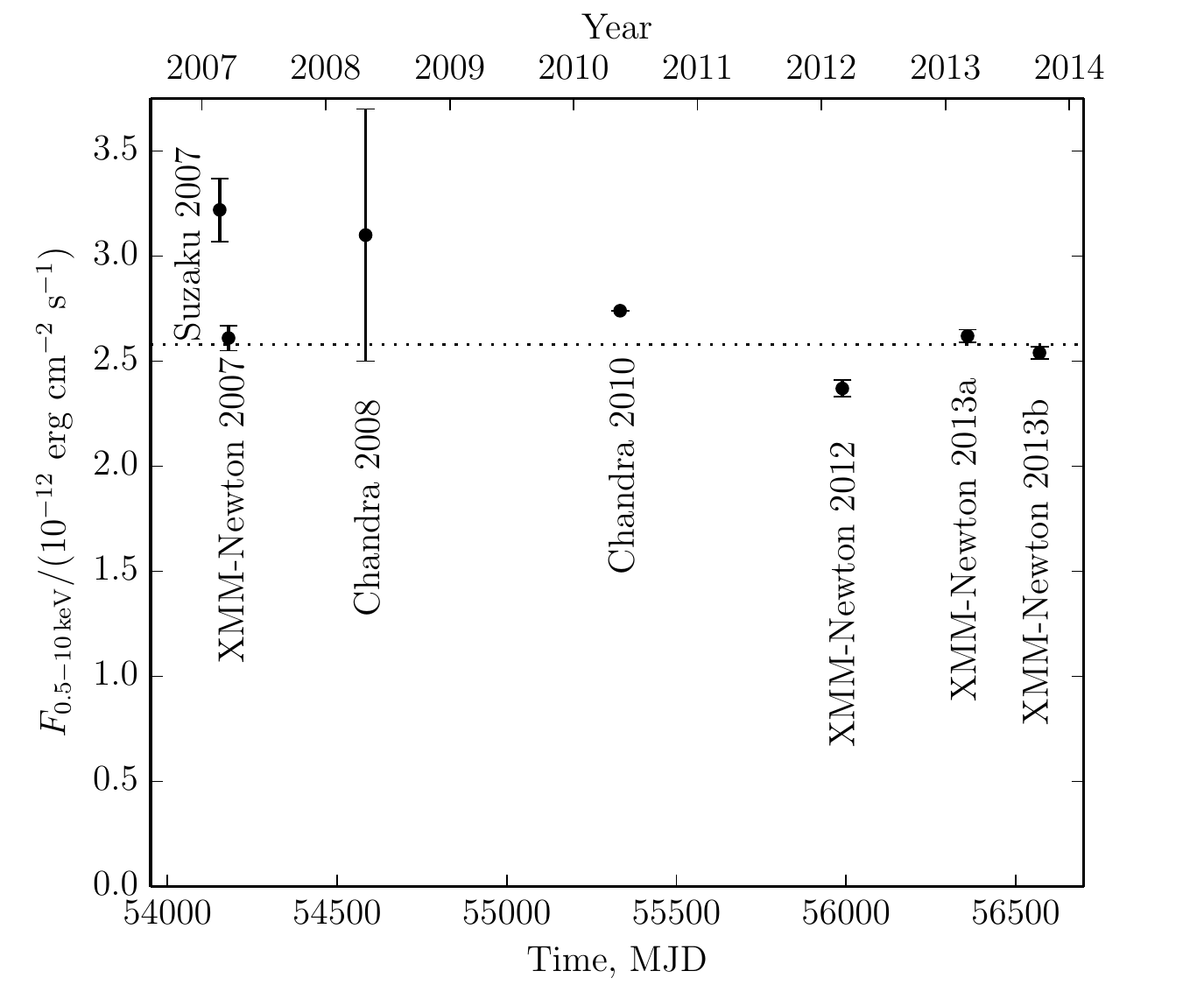}}
\caption{The absorbed flux measurements of the \cco in the
  0.5--10\,keV energy range performed with different instruments. 
  The error bars indicate only statistical errors. For
  the data point of 
  \emph{Chandra} 2010 taken from \citet{Halpern:Gotthelf:10c}, no error
  bar is available. 
  The latest observations analyzed in this work are
  marked as \emph{XMM-Newton 2013a} and \emph{XMM-Newton 2013b}.
  The horizontal dotted line indicates the averaged
  flux of the three \emph{XMM-Newton} observations (2007, 2013a, and
  2013b) performed in the same imaging mode.}
\label{fig:lclong}
\end{figure}

The new observations extend the long-term
monitoring of the source flux which now covers almost seven years.
Figure\,\ref{fig:lclong} shows the measured 
0.5--10\,keV flux (absorbed) as a function of time. 
The \emph{Chandra 2010} data point is taken from
\citet{Halpern:Gotthelf:10c}. The error bars indicate only statistical
errors. The differences between the
flux measurements with different instruments or with the same
instrument in different observing modes (imaging/timing) reach
$\sim$20\%. They can, however, be attributed to imperfect absolute
flux calibration of different instruments and
modes\footnote{\url{http://xmm2.esac.esa.int/docs/documents/CAL-TN-0052.ps.gz}}. 
The only data points in Fig.\,\ref{fig:lclong} which can be 
compared directly are the \emph{XMM-Newton} observations in 2007 and 2013: all
three observations have been performed in the imaging
\emph{Full Frame}\footnote{\url{http://xmm.esac.esa.int/external/xmm_user_support/documentation/uhb/epicmode.html}}
mode. The measured fluxes in these observations turn out to be consistent
within the statistical errors. The averaged flux over the three
observations (indicated by the horizontal dotted line) is
$2.58(4)\times 10^{-12}$~erg~cm$^{-2}$~s$^{-1}$.
Summarizing the results presented above, we conclude
that the available flux measurements are consistent with the source
flux being constant over the period covered by the observations. 

\section{Upper limits on the pulsed fraction\label{sec:uplim}}

The \emph{XMM-Newton} observations in \emph{imaging mode} analyzed here
permit a search for periodicity below the Nyquist frequency of 
$f_{\rm Nq} = 1/(2t_{\rm res}) = 6.87$\,Hz corresponding to the pulse
period of 0.147\,s ($t_{\rm res}=0.0734$\,s is the time resolution of
EPIC-PN camera in the imaging mode, \citealt{Strueder:etal:01}). Since the lowest and the highest
measured pulse periods of CCOs are $\sim$0.105 and 0.424\,s, there is
a good chance to miss pulsations in \xmmu because of the
timing resolution limit. We note, however, that the analysis of the $\sim$20\,ksec
timing observations with \emph{XMM-Newton} presented in \paper did
not reveal any pulsations with the pulsed fraction above $\sim$10\%
down to a pulse period of 0.2\,ms. 

We searched for pulsations in the EPIC-PN data of both
\emph{XMM-Newton} observations performed in 2013. We calculated a
Rayleigh periodogram \citep[$Z_1^2$-statistics, e. g.,][and references
therein]{Protheroe:etal:87} for the filtered EPIC-PN event 
files in the energy range 0.35--5.5\,keV extracted with the Science
Analysis System (SAS) \footnote{\url{http://xmm.esac.esa.int/sas/current/documentation/sas_concise.shtml}} version 13.5.0. The standard filtering including
the removal of high flaring background intervals has left $\sim$53.2
and $\sim$58.0\,ksec of useful source exposure in the March and
October observations,  
respectively. The energy range is a priori chosen as rough
compromise between the source-to-background ratio and the total amount of 
signal counts. No pulsations 
have been found above 99\% confidence level in either of the two observations. 
The maximum value of $Z_1^2$-power is 27.3 for the March and
28.9 for the October observations, respectively, whereas the
99\%-threshold values of $Z_1^2$ in the
respective observations are 34.8 and 35.0.
The threshold level in each case is calculated taking into account
the number of trials which is equal to the number of independent
frequencies in the periodogram, $\sim$3.6$\times 10^5$ and 
$\sim$4.0$\times 10^5$, in the March and October observations,
respectively.  

The maximum values of $Z_1^2$ are converted to upper limits on the
pulsed fraction of the source assuming sinusoidal pulsations and using the
distributions of $Z_1^2$ for the case where both the signal and the noise are
present in the data 
\citep[e.g.,][]{Jager:etal:89}.
For the March observations,
the upper limit on the pulsed fraction of the source signal is
7.3\%. For the October observations, it is 8.1\%.

\section{Spectral analysis\label{sec:spe}}

We extracted spectra of the \cco taken with all EPIC cameras 
\citep{Strueder:etal:01, Turner:etal:01} onboard
\emph{XMM-Newton} during the March and October observations
2013. As for the timing analysis above, we used the version 13.5.0 of SAS
and the latest calibration files available at the time of the data
reduction (February--March 2014). The periods of high flaring
background have been identified and excluded leading to the reduction
of the total exposure by $\sim$25--30\% for the 2013 observations.
To combine the new spectra with
those obtained with \emph{XMM-Newton} in 2007
(Table~\ref{tab:obs}, \paper) we re-extracted the 2007 data using the
same software and calibration which we used for the 2013 data. 
In each case, we extracted the source spectrum from a circle
with a radius of 40 arcsec centered at the CCO position and encompassing
$\sim$90\% of the PSF energy below $\sim$5\,keV. The background
spectra are extracted according to the recommendations of the
instrument team, that is from the regions on the same CCD chip having
roughly the same RAW~Y coordinate as the 
source for EPIC-PN and from the elliptical
annuli with the inner and outer minor semi-axes of 
$\sim$55$^{\prime\prime}$ and $\sim$85$^{\prime\prime}$ elongated
along the circle of equal off-set radii for EPIC-MOS. The regions with
visible diffuse emission and stray light are excluded. 
The independent spectral analysis of the the three
observations did not reveal any variation of the source's spectral shape. The
spectra, responses, and background files from the three observations
have thus been combined separately for EPIC-PN, EPIC-MOS1, and EPIC-MOS2
cameras using the SAS tool \texttt{epicspeccombine}. All spectral
modeling described in the following is performed using the combined spectra.

\subsection{Blackbody versus model atmosphere fits}
\label{sec:bbody}

We modeled the combined spectra from the three observations with
a blackbody function and, alternatively, using the
carbon\footnote{\url{http://heasarc.gsfc.nasa.gov/xanadu/xspec/models/carbatm.html}} 
\citep{Suleimanov:etal:14} and hydrogen stellar atmosphere models
developed by our group. A multiplicative component accounting for the
low-energy photo-electric absorption by the interstellar medium was
added in each case. The stellar atmosphere models are applied in the 
same way as described in \paper. 
Further details on the stellar atmosphere modeling are provided in the
following subsections.

\begin{figure}
\centering
\resizebox{\hsize}{!}{\includegraphics{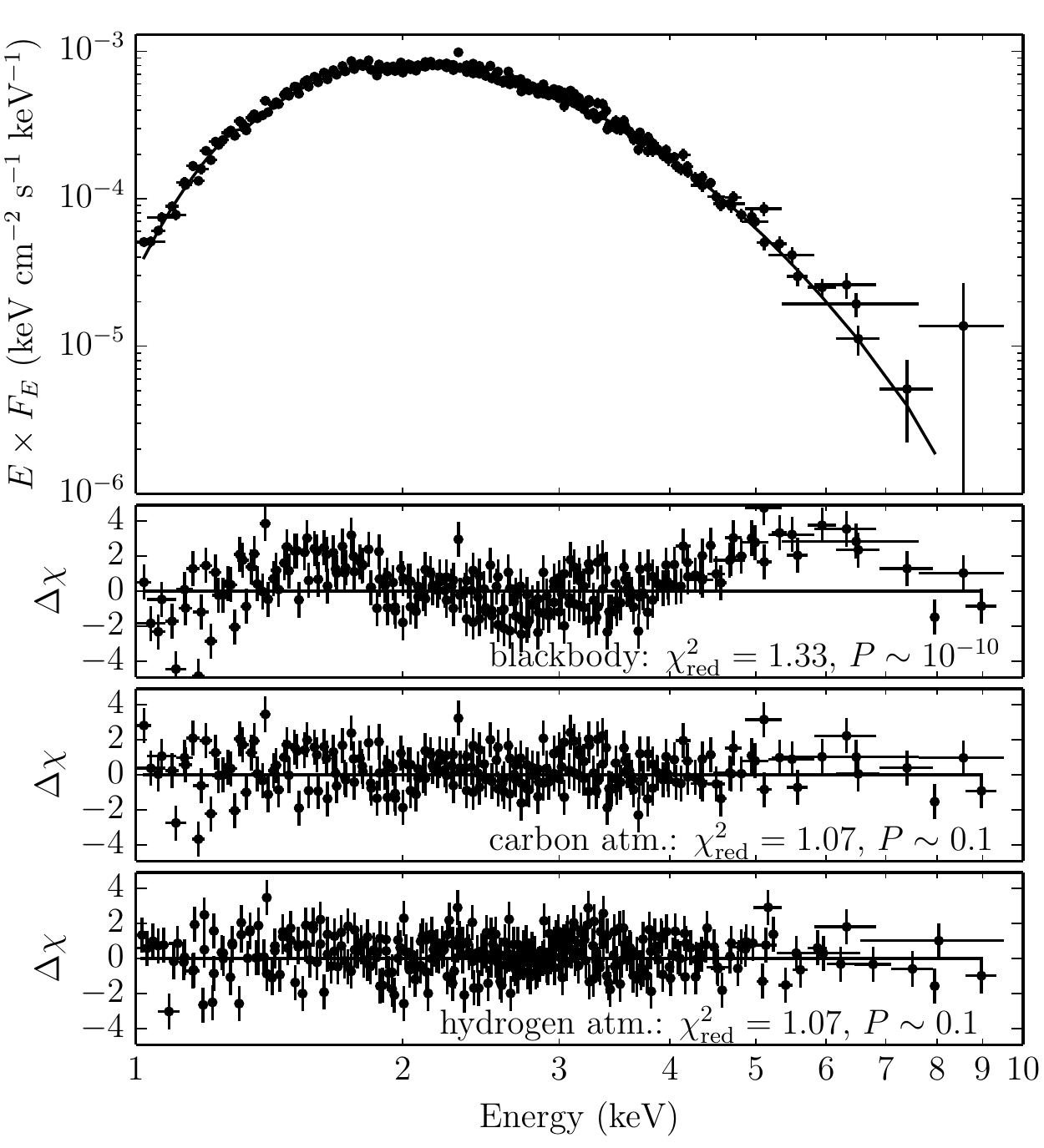}}
\caption{The combined \emph{XMM-Newton} spectrum (all EPIC cameras)
  from the three observations fitted with a carbon atmosphere model 
  (\emph{top}) and the fit residuals for the blackbody, the carbon
  atmosphere, and the hydrogen atmosphere models (from top to
  bottom). The null-hypothesis probabilities ($P$-values) yielded by
  the $\chi^2$-fits are indicated. The blackbody model leads to a
  statistically unacceptable fit.}
\label{fig:spe}
\end{figure}

The analyzed spectra allow for the first time the blackbody model to be
clearly rejected for this source. 
The corresponding fit yields a reduced
$\chi^2=1.33$ for 896 degrees of freedom (d.o.f.). This 
corresponds to a null-hypothesis probability ($P$-value) of 
$\lesssim$10$^{-10}$ indicating a statistically unacceptable fit. The
residuals of the fit show systematic wave-like structure (second from
top panel in Fig.\,\ref{fig:spe}). The best-fit parameters are the
following: 
$n_{\rm H}=(1.50\pm 0.02)\times 10^{22}$\,atom\,cm$^{-2}$, 
$kT=0.489\pm 0.002$\,keV.

The fits with the
carbon and hydrogen stellar atmosphere models yield flat residuals
(two bottom panels of Fig.\,\ref{fig:spe}) and acceptable fits with
a reduced $\chi^2=1.07$ for 894 d.o.f. and $P$-values of $\sim$0.1
in both cases. A more complicated spectral model
than a simple blackbody is thus required solely by the shape of the measured
spectral continuum. 
 
We verified that the spectral fits described above are not
sensitive to the exact choice of the background regions. We thus
believe that possible systematic variations of the background over
the detector plane close to the source position do not affect our
conclusion. We refer to  
Sect.\,\ref{sec:carbon} for a discussion of possible systematic effects
related to background.

\subsection{Fit with carbon atmosphere models}
\label{sec:carbon}

Following \paper, we assumed that the atmosphere characterized
by the effective temperature $T$ covers the entire surface of a compact
star with mass $M$ and radius $R$ located at a certain distance
$d$. Mass and radius are used to calculate the surface
gravity and the gravitational redshift affecting the atmosphere model
while the distance determines the normalization 
constant $K$ of the spectrum: $K = 1/d^2$. The mass,
radius, effective temperature, and the equivalent hydrogen column density
$n_{\rm H}$ which characterizes the photo-electric absorption by
the interstellar medium are free fit parameters of the spectral model. 

The gravitational redshift $z$ and the surface gravity
$g$ affect the shape of the model spectrum in
different manners \citep[see][for
details]{Suleimanov:etal:14}. Both values can thus be determined
independently from a spectral fit. On the other hand, $M$ and $R$
enter the formulae for $z$ and $g$ in different ways: 
$1+z=[1-2GM/(c^2R)]^{-1/2}$, $g=GM/R^2(1+z)$. That is, the curves of
equal $z$ in the $M-R$ plane are not parallel to the curves of
equal $g$ \citep[see, e.g., Fig.~2 of][]{Suleimanov:etal:14}.  A
certain pair of $g$ and 
$z$ values thus uniquely determines a point in the  $M-R$ plane.
Therefore both mass and radius can in principle be determined independently from
a spectral fit as was also demonstrated in \paper.

\begin{figure}
\centering
\resizebox{\hsize}{!}{\includegraphics{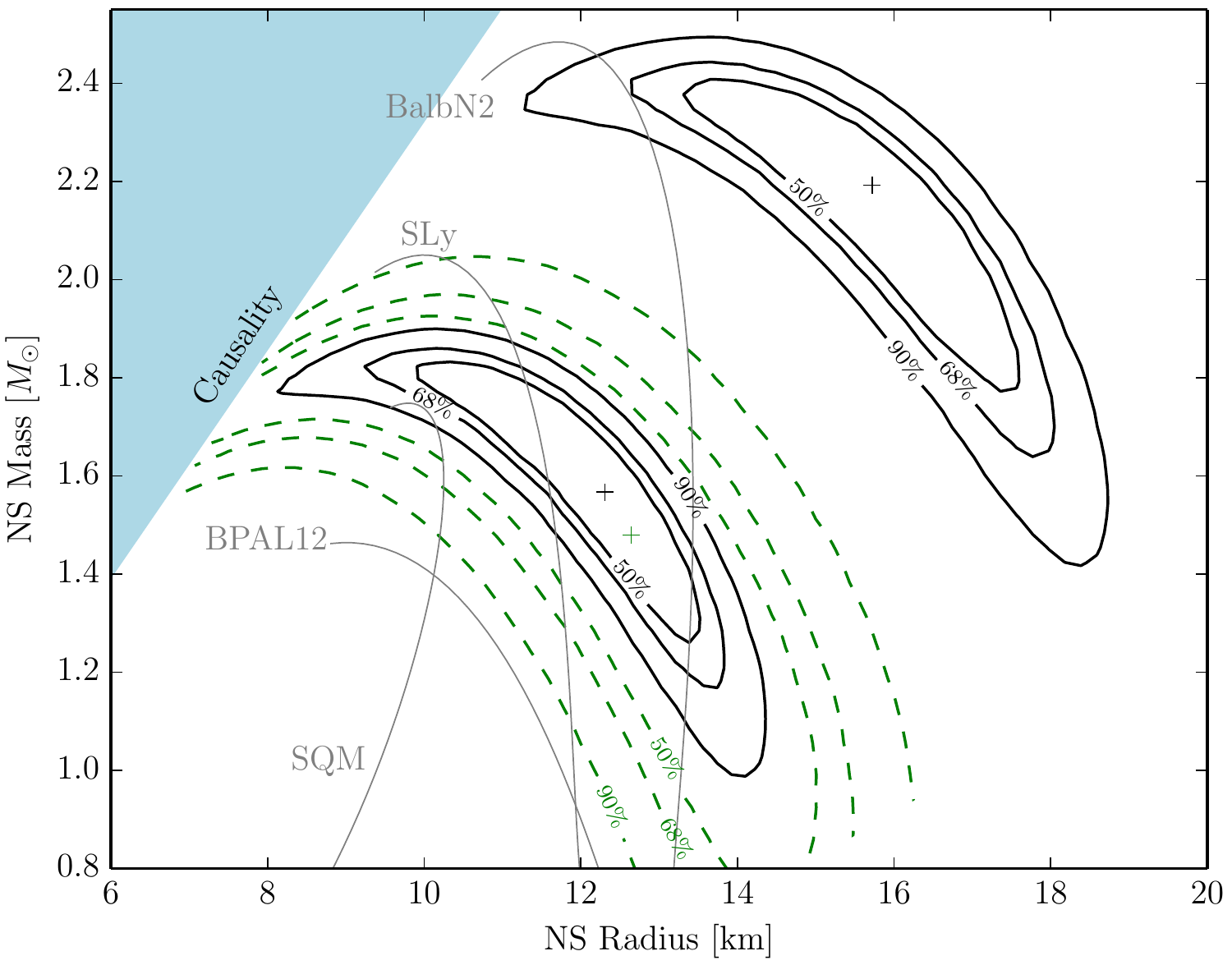}}
\caption{The $\chi^2$ confidence regions in the mass--radius
plane for the \cco obtained with the
carbon atmosphere models for the fixed distances of 3.2\,kpc (bottom
left) and 4.5\,kpc (top right). The indicated confidence levels (50,
68, and 90\%) are for \emph{two} parameters of interest.
The crosses indicate the corresponding
$\chi^2$-minima. The shaded area in the top left
of the plot indicates the region excluded by the requirements of
causality. The dashed contours indicate the confidence regions only with
the data from the 2007 observations \citep[presented
in][]{Klochkov:etal:13}. The thin curves indicate the mass--radius
dependencies for some of the commonly used nuclear EOS.} 
\label{fig:RMcarb}
\end{figure}

As in \paper, we first used two fixed distances to the source,
3.2 and 4.5\,kpc, corresponding to the location of the SNR in the
Scutum-Crux or Norma arms, respectively \citep{HESS:2011}.
The resulting confidence regions in the $M-R$ plane are presented in 
Fig.\,\ref{fig:RMcarb}. For $d=3.2$\,kpc we also show the
contours obtained with the 2007 observations only (presented in
\paper) for comparison. The dramatic improvement in the statistics
(reduction of the contours size) is caused by the increase in the
total exposure time by a factor of $\sim$five. 
For the best-fit parameters and the associated statistical
uncertainties (1\,$\sigma$ c.l. for \emph{one} parameter of interest) we refer
to Table~\ref{tab:spe}.
We note that the confidence levels of the contours in
Fig.\,\ref{fig:RMcarb} are plotted for \emph{two} parameters of
interest. The contours corresponding to 68\% (1\,$\sigma$)
confidence level for
\emph{one} parameter of interest (whose projections to the $M$- and
$R$ axes correspond to the uncertainties given in
Table~\ref{tab:spe}) would be located slightly within the innermost
contours shown in the plot.

\begin{figure}
\centering
\resizebox{\hsize}{!}{\includegraphics{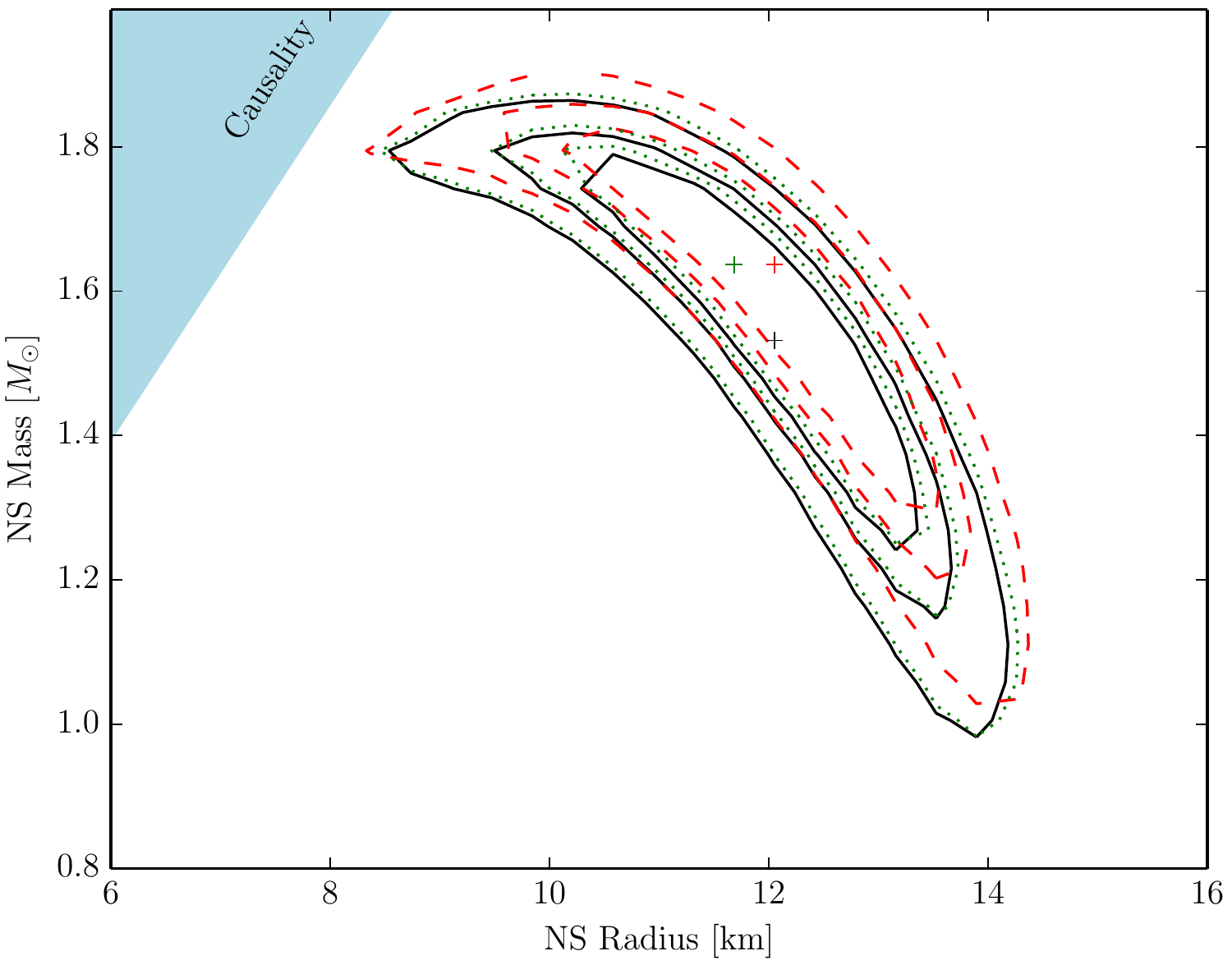}}
\caption{The confidence regions similar to those shown in
  Fig.\,\ref{fig:RMcarb} for the 
  fixed distance of 3.2\,kpc. The different colors and line styles
  indicate the contours obtained with three independent sets of background
  regions used for the corresponding background spectra extraction.} 
\label{fig:RMdiff}
\end{figure}

The presented contours take into account only statistical uncertainties
of the source and background spectra. Additional systematic
effects may be related to the spatial variation of the
background. Although the background regions are selected close to the
source to represent the ``local'' background (see above), they might
still include contributions from the diffuse emission and stray light
not directly visible in the image or hidden behind the source's PSF as well as the variations of the
instrumental background with respect to that at the source position. To assess
possible systematic uncertainties related to these effects, we
produced three independent sets of local background regions.
Figure~\ref{fig:RMdiff} presents the confidence contours for the
fixed distance of 3.2\,kpc (similar to those shown in the bottom left of
Fig.\,\ref{fig:RMcarb}) obtained with the three sets of background
regions. The relative displacements of the contours shown with different
colors and line styles reflect the aforementioned systematics. 
One can see that the displacements
are by a factor of several smaller than the characteristic size of the
contours. We thus conclude that the
investigated systematics can be neglected at this stage.

\begin{table}
  \centering
  \renewcommand{\arraystretch}{1.3}
  \caption{Results of the spectral modeling of the \cco with
    carbon atmosphere models. The indicated uncertainties are at
    1\,$\sigma$ c.l. for one parameter of interest.}
  \label{tab:spe}
  \begin{tabular}{l l l}
    \hline\hline
  $d$, kpc (fixed)         &    3.2        &   4.5     \\
    \hline
                             &               &          \\
$n_{\rm H}/(10^{22}$\,cm$^{-2})$&$2.00\pm 0.03$&$1.99_{-0.03}^{+0.02}$\\
 $T$, MK                &$2.24_{-0.13}^{+0.39}$&$2.43_{-0.23}^{+0.22}$\\
 $T_\infty=T/(1+z)$, MK  &$1.78_{-0.02}^{+0.04}$&$1.82_{-0.01}^{+0.03}$\\
 $M/M_\odot$             &$1.55_{-0.24}^{+0.28}$&$2.19_{-0.34}^{+0.19}$\\
 $R$, km                &$12.4_{-2.2}^{+0.9}$&$15.7_{-2.0}^{+1.6}$\\
 $\chi^2_{\rm red}$/d.o.f.     & 1.07/895      &  1.07/895 \\
    \hline
  \end{tabular}
\end{table}

\begin{figure}
\centering
\resizebox{\hsize}{!}{\includegraphics{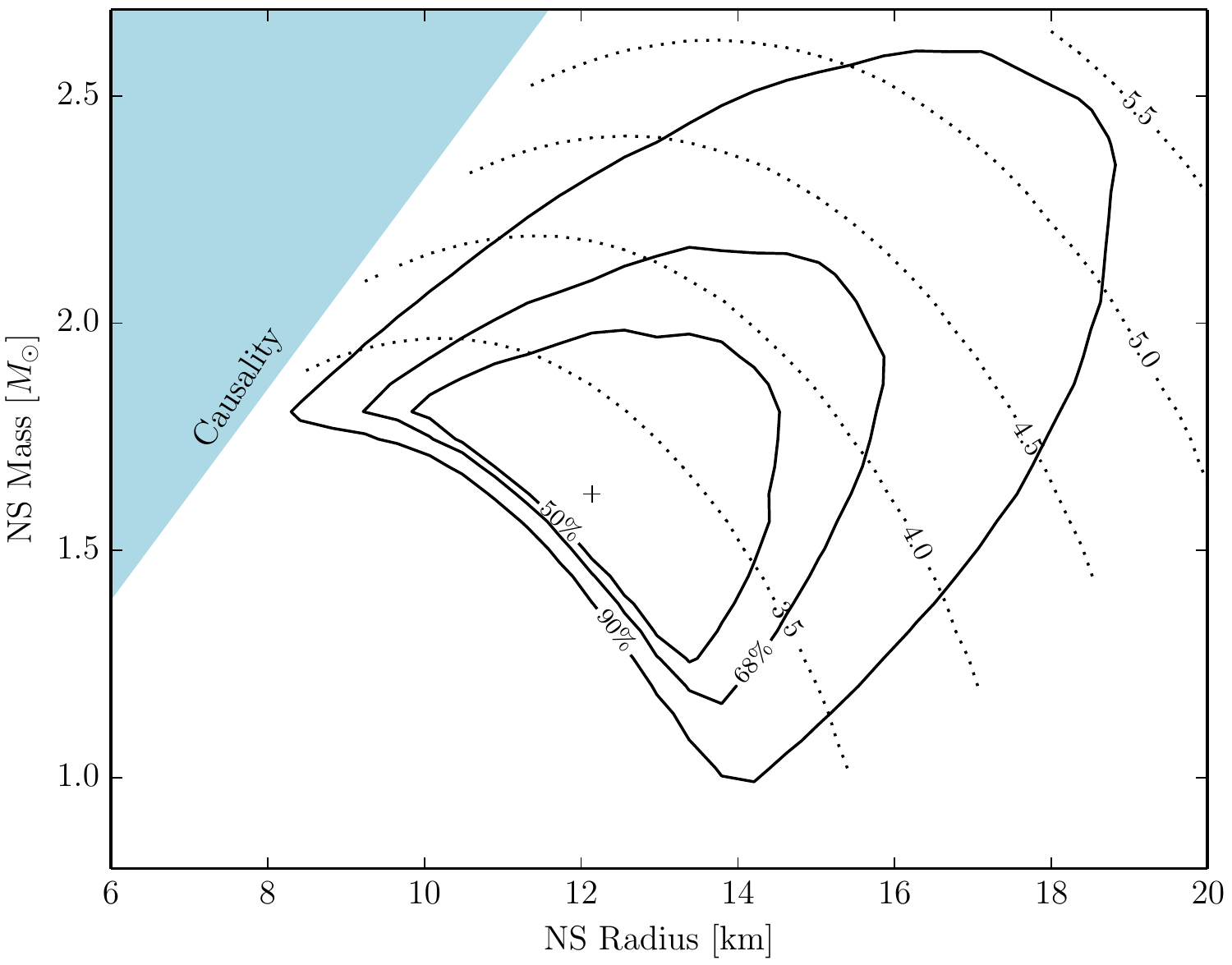}}
\caption{The confidence regions similar to those shown in
  Fig.\,\ref{fig:RMcarb} but with the distance as a free fit parameter
with a lower limit of 3.2\,kpc. The dotted contours indicate best-fit
distance in kpc computed for each point of the mass-radius grid.} 
\label{fig:RMd}
\end{figure}

As can be seen in Table~\ref{tab:spe},
the spectral models with $d=3.2$\,kpc
and $d=4.5$\,kpc provide the same fit quality. Nevertheless, the model
is sensitive to the distance as seen in
Fig.\,\ref{fig:RMd}. The figure shows the $M-R$ contours with the distance
being a free fit parameter with a lower limit of 3.2\,kpc. The
best-fit distances are computed for each point of the $M-R$ grid and
are indicated with the dashed contours. 
Distances larger than $\sim$5.5\,kpc lead to high $\chi^2$ values and
are formally excluded under the assumption that the   
emitting region is the entire stellar surface.

\subsection{Fit with hydrogen atmosphere models}
\label{sec:hydrogen}

The combined \emph{XMM-Newton} spectra of \xmmu were also fitted using
our hydrogen NS atmosphere models. The model spectra are calculated
for a range of effective temperatures of 0.5--10\,MK with a step size of
0.05\,MK and for the same set of nine surface gravity logarithms $\log
g$ as for our carbon atmosphere models \citep{Suleimanov:etal:14}.  The
models are computed using the LTE approximation with pressure
ionization effects included using the occupation probability
formalism \citep{Hummer:Mihalas:88} as described by \citet{Hubeny:etal:94}.  
The method of modeling is described in detail by \citet{Suleimanov:Werner:07}.
We do not consider Compton scattering; all models are computed using
coherent electron scattering alone. This approach is not correct for hot model
atmospheres with $T > 3-5$\,MK
\citep{Suleimanov:Werner:07}. Therefore, the hottest 
models have to be used with caution. 
It turns out, however, that the fits to our spectra yield
lower temperatures.
Similar to the carbon atmosphere models (previous section),
the free parameters of the spectral model are the mass and radius
of the star, the effective temperature $T$, the equivalent hydrogen
column density, and the distance $d$.

\begin{figure}
\centering
\resizebox{\hsize}{!}{\includegraphics{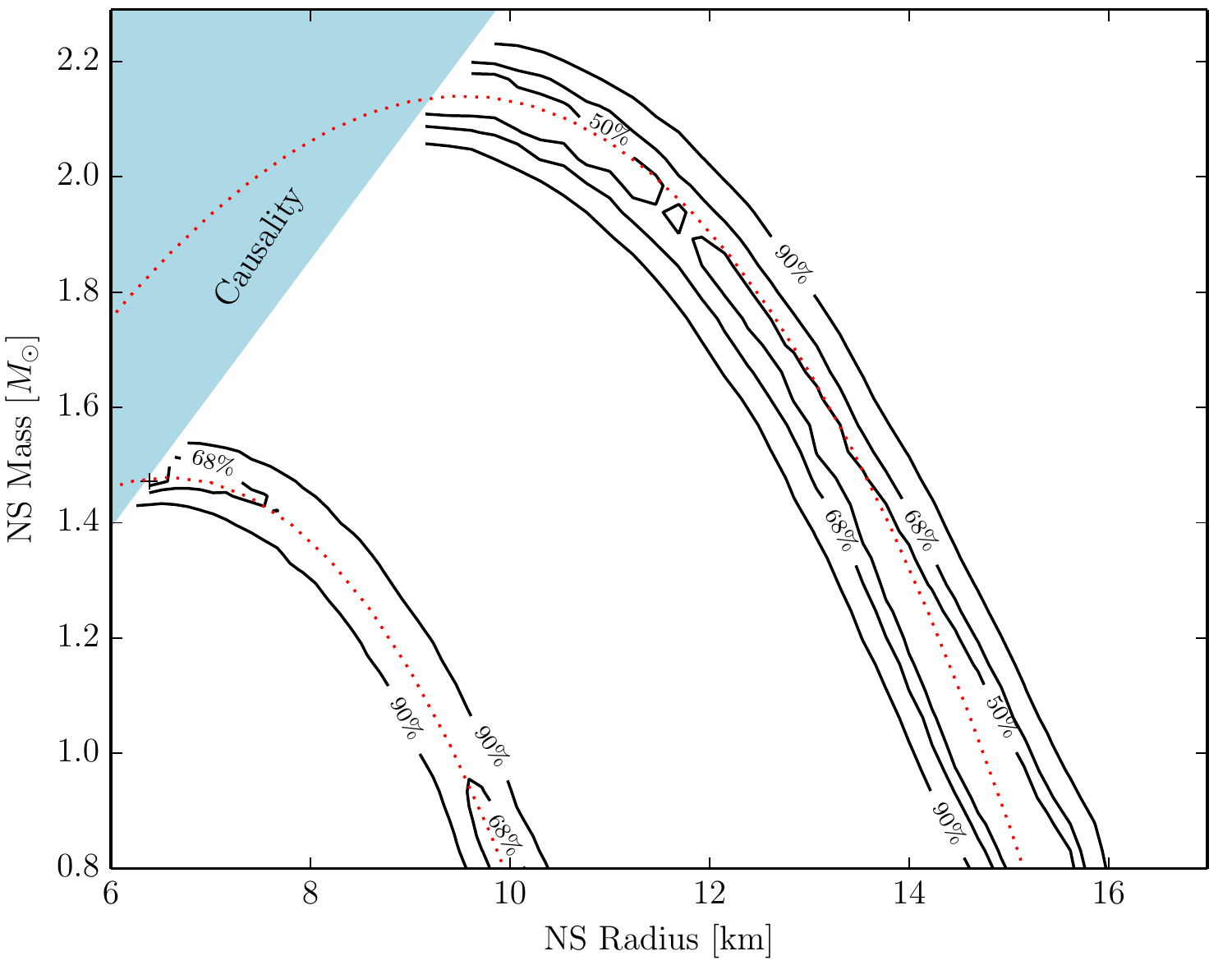}}
\caption{The confidence regions similar to those in
  Fig.\,\ref{fig:RMcarb} but for the hydrogen atmosphere models and for
  the fixed distances of 7\,kpc (bottom left) and
  10\,kpc (top right). The dotted curves indicate the lines of equal
  apparent radii: $R_\infty=11.4$\,km (bottom) and $R_\infty=16.5$\,km 
  (top). The corresponding best-fit apparent temperatures are 
  $2.76\pm 0.02$ and $2.78\pm 0.02$\,MK for 7 and 10\,kpc,
  respectively. 
}
\label{fig:RMHatm}
\end{figure}

As mentioned in Sect.~\ref{sec:bbody}, the fit quality in case
of the hydrogen models is similar to the carbon model: 
$\chi^2_{\rm red}\simeq 1.07$ for 895 d.o.f. (the corresponding $P$-value is
$\sim$0.1).
The confidence contours in the $M-R$ plane resulting from the fit with
the hydrogen models are presented in Fig.\,\ref{fig:RMHatm} for two
fixed distances, 7 and 10\,kpc. The distances were chosen for
illustration only. For $d<7$\,kpc the contours require masses
and radii smaller than the canonical values usually assumed for an NS.
The distances above 10\,kpc would imply an unrealistically high TeV
luminosity of the SNR. Thus, under the assumption that the emission
is uniformly produced by the entire stellar surface, a hydrogen
atmosphere seems to be incompatible with the canonical masses and
radii of an NS and with the available constraints on the source
distance. 

It is interesting to note that the contours are located along
the respective lines of equal apparent radii indicated by the dotted
curves. The model thus behaves similar to a blackbody function
where the apparent radius is coupled to the normalization for a fixed
distance and is therefore a fit parameter. The apparent temperatures
$T_\infty = T/(1+z)$ measured with the hydrogen models are 
$2.76\pm 0.02$ and $2.78\pm 0.02$\,MK for 7 and 10\,kpc, respectively.

For completeness, we provide the best-fit absorption column density
obtained with the hydrogen model, $n_{\rm H}=(1.91\pm 0.03)\times
10^{22}$\,atom~cm$^{-2}$. Since the confidence contours are not closed
(Fig.\,\ref{fig:RMHatm}), mass, radius, and effective
temperature are formally unconstrained by the fit. 

As mentioned in \citet{Suleimanov:etal:14} and \paper, a
number of effects are not taken into account  in our calculations of the
carbon atmosphere models such as
the influence of the magnetic field and a possible complex chemical
composition different from pure carbon. The same applies to
the hydrogen atmosphere models described above.  These effects might
influence the shape of the spectrum and thus the presented $M-R$
confidence contours. We believe, however, that our models are
applicable for $B\lesssim 10^8-10^9$\,G as explained in
\paper. Since no pulsations are detected in \xmmu,
no estimates of the surface $B$-field strength are available and our
assumption of low magnetic field can be valid.

We also note that the values of the interstellar absorption column
density obtained with both carbon and hydrogen models 
are within the range of the column densities measured in
different parts of the remnant \citep{HESS:2011}.

\section{An exceptionally hot cooling neutron star\label{sec:cooling}} 

\begin{figure}
   \resizebox{\hsize}{!}{\includegraphics{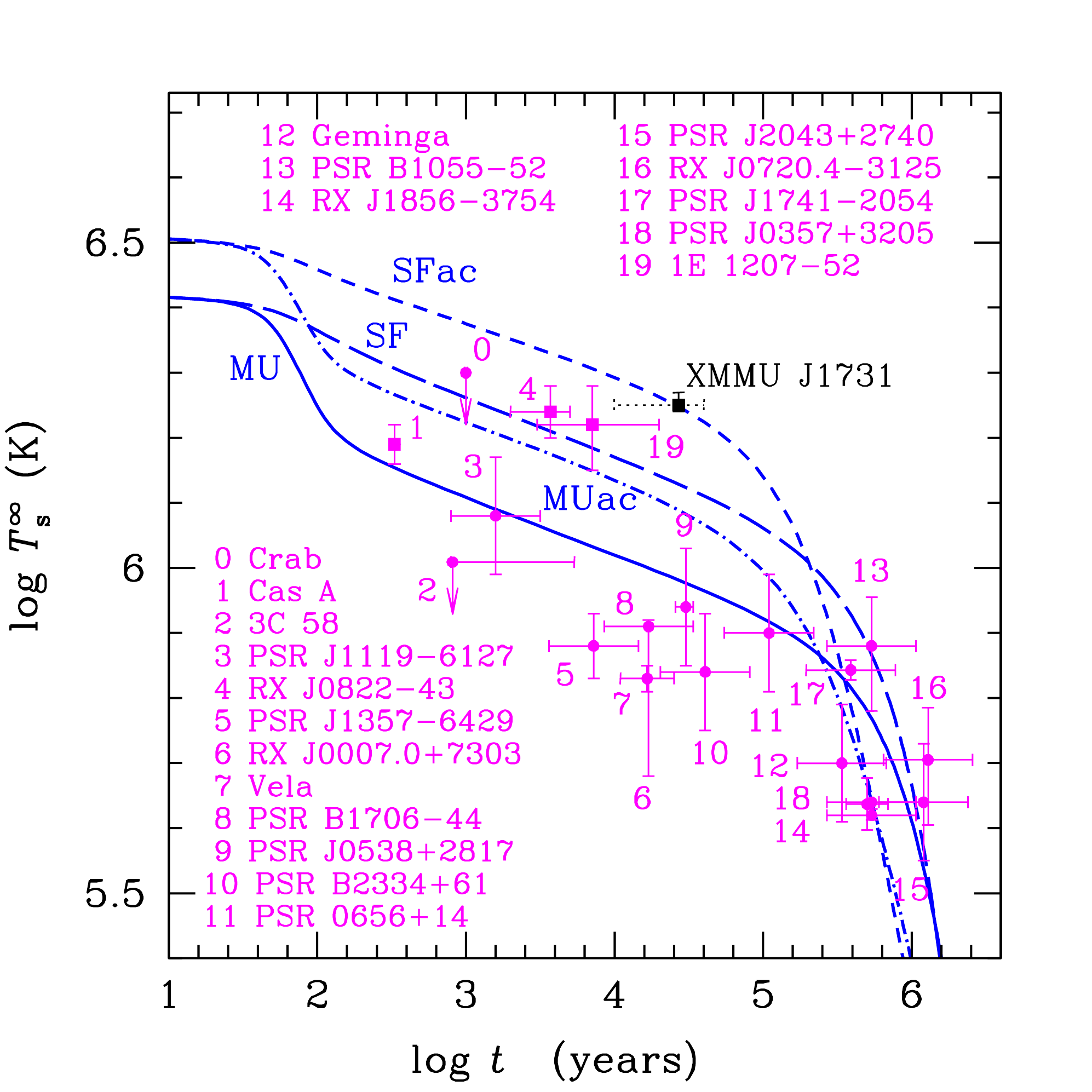}}
   \caption{The effective surface temperatures or upper limits for a
     number of cooling
	isolated NSs including \xmmu versus their ages
        (data points) compared 
	with four theoretical cooling curves for a 1.5\,${\rm
          M_\odot}$ star.
        The plot contains four CCOs with reliable temperature
          measurements (labeled with squares). 
        For \xmmu, the dashed error bar indicates a conservative range
        of 10--40\,kyr adopted for the NS age.
	MU refers to a non-superfluid star with a heat
        blanket made of iron; 
	SF corresponds to a star with strong proton superfluidity in the core and
	the same heat blanket; MUac and SFac refer, respectively, 
	to the same models as MU and SF but with a fully carbon heat
        blanketing envelope (see text for details).}
   \label{hesscool}
 \end{figure}

Since the apparent temperature of the \cco is well measured,
it is interesting to investigate its status among other NSs of this
family. Figure \ref{hesscool} shows a representative sample of cooling
isolated NSs in a $T_s^\infty-t$ diagram, where
$t$ is the estimated/measured age of the objects and $T_s^\infty$ is the
apparent effective surface temperature (here we denote the surface temperature
as $T_s$ to distinguish it from the internal temperature of the star,
$T_i$). The data are the same as in \citet{Weisskopf:etal:11} with an
addition of three new sources,
\xmmu (this work), PSR J1741--2054 \citep{Karpova:etal:14}, and
PSR\,J0357+3205 \citep{Marelli:etal:13,Kirichenko:etal:14}. We have also added
1E\,1207.4--5209 \citep{Zavlin:etal:98}.
 
The data displayed in Fig.\,\ref{hesscool} contain four CCOs which  
are marked by squares (to distinguish from other sources 
labeled by filled dots). Thermal emission of two CCOs,
the NS in Cas\,A and \xmmu, can be interpreted as
radiation from the entire star surface with realistic mass and
radius using the carbon atmosphere models as explained in detail
in \citet{Ho:Heinke:09}, \paper, and in this work.
As for RX\,J0822--4300 (CCO in the SNR Puppis\,A) and 1E\,1207.4--5209 (CCO
in the SNR G296.5+10.0), their emission can be 
interpreted with hydrogen atmosphere models 
\citep{Zavlin:etal:98,Zavlin:etal:99}.

The indicated position of \xmmu corresponds
to $d=3.2$\,kpc and $t=27$\,kyr. The age was estimated by \citet{Tian:etal:08} 
using the Sedov solution for the SNR
assuming a distance of 3.2\,kpc. 
The error bars on $T_s^\infty$ are enlarged by a factor of two
for a better visualization.
The uncertainties of the age estimate are not
well understood. We adopt a conservative range of
10--40\,kyr. For larger distances, the age must be larger. 
One can see that the CCO appears exceptionally hot for
the assumed age, i.e., it must have been cooling very slowly
posing a challenge for the cooling theory.

According to the 
theory, middle-aged isolated NSs (as those shown in Fig.\ \ref{hesscool})
have already passed the initial cooling stage of internal thermal 
relaxation -- the first 10--100 yrs depending on the NS model
\citep[e. g.,][]{Yakovlev:Pethick:04,Page:etal:09}. Their internal
layers are almost isothermal. A strong temperature gradient still persists in 
the outer heat blanketing envelopes (under the atmosphere, a few ten meters
thick). Such stars are mostly cooling ``from inside'' (mostly from their 
superdense cores) via neutrino emission. The photon thermal luminosity
only becomes significant in cooling at $t \gtrsim 10^5$\,yr. 
The decay of the internal
temperature $T_i$ is determined by the neutrino cooling function
\begin{equation}
   \ell(T_i)=L_\nu(T_i)/C(T_i),
\label{e:fl}
\end{equation}
where $L_\nu(T_i)$ and $C(T_i)$ are the neutrino luminosity and the heat capacity
of the star, respectively.
Accounting for General Relativity, $T_i$, $L_\nu$, and
and $C$ have to be redshifted in the cooling equations.
Thus, the internal
cooling is regulated by the neutrino emission and the heat capacity.
The effective surface temperature $T_s$ simply adjusts itself to the 
internal temperature $T_i$. The adjustment is taken into account
using calculated $T_s(T_i)$-relations. These relations are determined
by the thermal conductivity  
of the heat blanketing envelopes which is affected by their chemical
composition.

The four theoretical cooling curves $T_s^\infty(t)$ presented in
Fig.\,\ref{hesscool} are calculated using 
a generally relativistic cooling code \citep{Gnedin:etal:01}. 
The curves are aimed at
explaining the hottest cooling NSs. 
Colder NSs are commonly treated as sufficiently massive stars with higher
neutrino emission from the core. In Fig.\ \ref{hesscool},  we use
an NS model with a nucleon core (the modification of
the APR EOS of superdense matter used in \citealt{Weisskopf:etal:11}).
The NS mass is
$M=1.5\,{\rm M_\odot}$ and radius $R=12.03$\,km. The cooling curves
are, however, known  to be fairly independent of EOS, $M$, and $R$ 
for typical theoretical neutron stars ($M \lesssim 1.8\,{\rm
  M_\odot}$, $R \sim 10-13$ km) where the direct Urca process does not
operate \citep[e.g.,][]{Yakovlev:Pethick:04}. 

The neutrino emission from the core of the given star
can be produced by the modified Urca process (MU) and somewhat weaker
processes of nucleon-nucleon collisions (but no powerful direct Urca process).
Neutron superfluidity in the core due to triplet-state pairing is 
neglected, proton superfluidity due to single-state pairing is varied.

Further we assume a blanketing envelope made of carbon with a mass of
$\Delta M$ \citep[][]{Yakovlev:etal:11}. 
For $\Delta M \lesssim 10^{-14}\,{\rm M_\odot}$, the amount of carbon
is too small to affect the cooling. 
At $\Delta M \sim 10^{-8}\,{\rm M_\odot}$ the amount of carbon is
maximal. For formally higher $\Delta M$, carbon would 
transform into heavier elements at the bottom of the envelope because of
beta captures and pycnonuclear 
reactions. We neglect the effects of the magnetic field on the cooling
which is a good approximation as long as $B \lesssim 10^{12}$~G.   

The solid cooling curve (MU) corresponds to a non-superfluid star with an
ordinary (iron) heat blanketing envelope. This is a basic cooling curve 
(cooling through the modified Urca process non-affected by
superfluidity, the so-called {\em standard neutrino candle},
\citealt{Yakovlev:etal:11}). It is consistent 
with observations of some NSs, but cannot
explain \xmmu which is substantially hotter.
One can obtain a hotter star by switching on proton superfluidity and replacing 
iron with carbon in the heat blanketing envelope.  
The long-dashed cooling curve (SF) corresponds to strong proton superfluidity
in the NS core (with the critical temperature over the core
$T_{\rm cp}(\rho)\gtrsim 3 \times 10^9$~K, where
$\rho$ is the density) and an iron heat blanket. 
The exact $T_{\rm cp}(\rho)$ profile in this limit is not important. 
Such superfluidity almost completely switches off the proton heat capacity 
(reducing the total heat capacity by $\sim$25\%, \citealt{Page:93})
and all neutrino processes involving 
protons, first of all the modified Urca process. The neutrino luminosity
from the core is determined by neutron-neutron collisions and is about two
orders of magnitude weaker than the modified Urca. This greatly
reduces the neutrino luminosity
function (\ref{e:fl}), slows down the internal neutrino cooling, and makes
the star hotter (at $t \gtrsim 100$ yr, after the internal thermal relaxation 
for the given NS model). However, it is still insufficient to 
explain the temperature of \xmmu at the assumed age.   

One might think that one can additionally slow down the cooling by assuming
strong superfluidity of neutrons in the core. It will indeed block
neutrino emission due to neutron-neutron collisions leaving a
much weaker neutrino emission caused by electron-electron
collisions. It will, however, also block the neutron heat 
capacity in the core (the main source of heat capacity there, \citealt{Page:93})
producing a 
neutron star with very low neutrino luminosity
and heat capacity in the core. Its cooling will be greatly affected by
the microphysics of the crust and will not be much slower
than that without neutron superfluidity
\citep[see, e.g.,][]{Yakovlev:Pethick:04}.

The dotted-dashed cooling curve (MUac) in Fig.\,\ref{hesscool} refers to
a non-superfluid star but with a fully carbon heat blanketing envelope
($\Delta M/M= 10^{-8}$). Internally, this star cools down as the
standard candle (through the modified Urca process) but the heat
blanket is now made of lighter, more heat transparent carbon. 
This increases the surface temperature \citep{Yakovlev:etal:11}. 
However, such an increase is still insufficient to explain \xmmu.

Finally, the short-dashed curve in Fig.\,\ref{hesscool} corresponds to
strong proton superfluidity and a fully carbon heat blanket. Now we
have used all cooling regulators
to slow down the cooling and to obtain
an exceptionally hot NS. This 
scenario is formally consistent with the data of \xmmu. 
Therefore, with the assumed age and distance, the data formally
require both very low neutrino cooling rate (low $\ell$) and a
fully carbon heat blanket. Such a slowly cooling NS would be
a remarkable object. The neutrino emission 
from its core is so slow that the neutrino emission from the crust 
starts to affect the cooling.

We repeat that the above considerations assume $t=27$\,kyr and
$d=3.2$\,kpc. For a distance of 4.5\,kpc, the Sedov
solution would yield  $t\simeq 40$\,kyr and the problems related to
the NS cooling would become even more severe.
The temperatures derived using the blackbody and
hydrogen models, which are not well justified 
(Sects.\,\ref{sec:bbody} and \ref{sec:hydrogen}),
turn out to be much higher than for the carbon
model so that the NS would appear even hotter.
If, however, \xmmu is younger, e.g., $t \sim 10$ kyr, the discrepancy 
with the cooling theory is reduced. However, even in that case we
would need both
strong proton superfluidity and a lot of carbon in the heat blanket, although 
superfluidity could be somewhat weaker (lower $T_{\rm cp}$) and/or
the amount of carbon could be lower (smaller $\Delta M$).

\begin{figure}
\centering
\resizebox{\hsize}{!}{\includegraphics{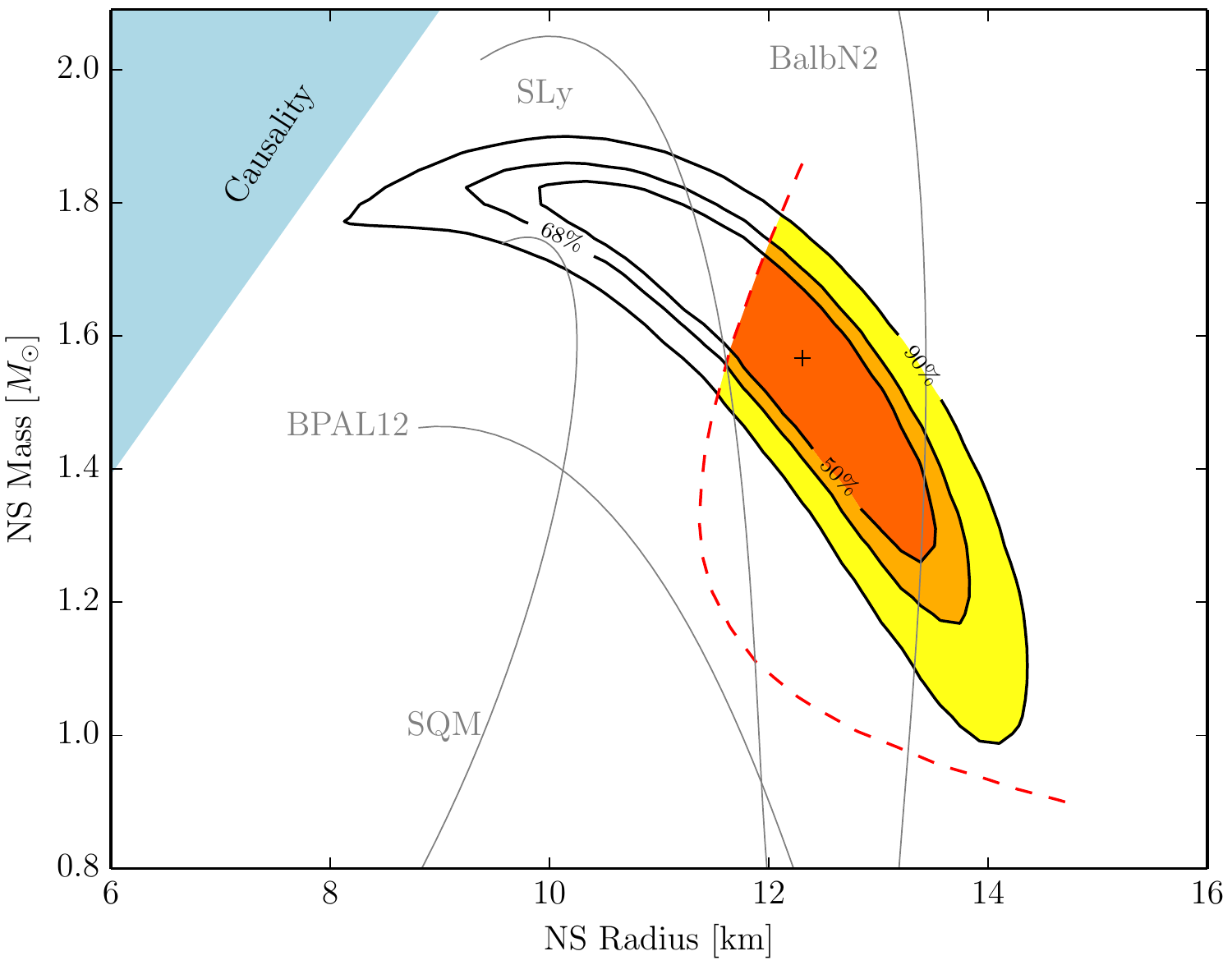}}
\caption{The confidence regions similar to those in
  Fig.\,\ref{fig:RMcarb} for a distance of 3.2\,kpc 
  where the area on the right side of the red dashed curve indicates
  the region allowed by the cooling theory assuming the NS age of
  27\,kyr (see text for details).
}
\label{fig:RMcooling}
\end{figure}

The case of $t=27$\,kyr requires extreme parameters of both cooling regulators
(strong proton superfluidity and maximum amount of carbon). 
As discussed above, the theoretical cooling curves are almost
independent of EOS, $M$ and $R$ in a broad range of $M$ and
$R$. However, they do depend on $M$ and $R$ outside that range. In
case of the observed $T_s^\infty$, they slightly depend on $M$ and
$R$. Therefore, the explanation of the observed $T_s^\infty$
can fail at certain $M$ and $R$ where the theoretical cooling curve goes below
the inferred $T_s^\infty$ even at the extreme parameters of the cooling
regulators. These values of $M$ and $R$ can be treated as
\emph{forbidden} by the cooling theory which gives
additional constraints on $M$ and $R$. 
A precise derivation of such constraints is beyond the scope
of this work and will be presented elsewhere. In
Fig.\,\ref{fig:RMcooling}, we present our preliminary constraints
for $t=27$ kyr. They are obtained using the rescaling relations which allows
us to calculate $T_s^\infty$ for a slowly cooling middle-aged NS in
a wide range of $t$, $M$, $R$, and EOSs. The relations are then 
calibrated with respect to theoretical cooling curves in Fig.\ \ref{hesscool}.
The allowed region in Fig.\,\ref{fig:RMcooling} is on the right side of the
red dashed curve. One can see that under the stated assumptions 
the cooling theory mostly eliminates the 
range of small $R \lesssim 12$\,km. 

As an alternative explanation of the exceptionally high $T_s^\infty$,
additional heating mechanisms inside the star 
can be invoked
\citep[e. g.,][]{Yakovlev:Pethick:04,Page:etal:06}.
In our case, these mechanisms are, however, not easy
to justify because they are known
to be mostly efficient in old and cold NSs whereas
\xmmu is very hot. 


\section{Discussion and conclusions\label{sec:discussion}}

The new observations continue the long-term monitoring of \xmmu and
confirm the lack of long-term variations in the source flux or
spectrum. Although the lack of any long-term variability is
considered to be a common property of CCOs, a secular
decrease of the NS effective temperature 
(``cooling'') was reported for the CCO in Cas~A by
\citet{Heinke:Ho:10}. The cooling is, however, questioned in the
recent work of \citet{Posselt:etal:13} where the authors
analyzed two \emph{Chandra} observations obtained in a pileup-free
observing mode.

The factor of five increase in the exposure time compared to the
previously available \emph{XMM-Newton} observations of \xmmu 
reported in \paper
substantially improves constraints on the spectral continuum of the
source. For the first time, the pure blackbody spectral function is
clearly rejected based solely on the fit quality rather than on the
best-fit values of the model parameters (as discussed, e. g., in \paper).
The carbon or hydrogen atmosphere models are clearly preferred by the
fit. 

The new observations did not reveal any pulsations with pulsed
fraction above $\sim$7--8\% and a period longer than $\sim$0.15\,s. In
combination with 
the absence of pulsations at higher frequencies reported in \paper 
with a similar upper limit on the pulsed fraction, this result supports
the hypothesis that the emitting area is identical or close to the
entire stellar surface. We maintain this hypothesis in the following
discussion. We admit, however, that the available upper limits on the
pulsed fraction of the order of $\sim$0.1 is not very stringent. Two
of the three known pulsating CCOs, RX\,J0822.0$-$4300 and
1E\,1207.4$-$5209, exhibit pulsed fractions of only 11 and 9\%,
respectively \citep[][and references therein]{Gotthelf:etal:13}.
One thus cannot completely reject the possibility that the emitting area
is somewhat smaller than the stellar radius. Stronger upper
limits on the pulsed fraction are necessary. 

Under the assumptions of the two most probable distances, 3.2 or
4.5\,kpc, the carbon atmosphere models lead to relatively compact
confidence regions in the NS mass--radius plane
(Fig.\,\ref{fig:RMcarb}). The contours corresponding to $d=3.2$\,kpc
are compatible with the most commonly used nuclear EOSs
and with the canonical NS
mass and radius of $\sim$1.4$M_\odot$ and $\sim$12\,km, respectively. 
The contours corresponding to $d=4.5$\,kpc require mass
and radius which appear to be uncomfortably large for an NS. 
The fit with the carbon atmosphere models thus supports
the possibility of the source located in
the Scutum--Crux spiral arm at $\sim$3\,kpc over the
location in the Norma arm at $\sim$4.5\,kpc. Distances
substantially larger than $\sim$5--6\,kpc are excluded solely by the
spectral fit under the stated assumptions (Fig.\,\ref{fig:RMd}). 
The sensitivity to the distance is a promising feature of the carbon
model which can be utilized in the the spectral analyses of isolated
low-magnetized NSs in the future. 

Interesting results concerning the applicability of
our carbon atmosphere models has recently been presented by
\citet{Bogdanov:14} who investigated the CCO in the supernova remnant
Kes~79. Contrary to all other CCOs, this object exhibits  
an extremely high pulsed fraction of $\sim$64\%.
The author demonstrates that the
X-ray spectrum of the source can be well fitted with the model of a pure
carbon atmosphere uniformly covering the entire
surface of a star with a radius of $\sim$14\,km at the adopted distance of
7\,kpc although such a uniformity apparently contradicts the observed high
pulsed fraction. Therefore, further searches for pulsations or the
improvements of the
upper limits on the pulsed fraction
for the CCOs in Cas\,A and in \hessj are of key importance for the
chemical composition of the NS atmosphere and the derived
parameters of the NSs.

The hydrogen atmosphere models are applied to \xmmu for the first
time. Contrary to carbon, they turn out to be not very sensitive to the
distance. The mass-radius contours obtained for a fixed distance are
not closed. For distances below $\sim$7\,kpc the contours
indicate the masses and radii which are substantially lower than those
usually assumed for NSs. However, they are compatible 
with a possibility of a \emph{strange} star covered by ordinary matter
(hydrogen). Such hypothetic stars built entirely or predominantly
of self-bound quark matter may possess masses and radii much smaller
than those of canonical NSs
\citep[e.g.,][]{Haensel:etal:07}. Besides this possibility and under
the assumption that the emitting area is the entire star surface, the
hydrogen atmosphere model seems to be incompatible with the available
distance estimates of $\sim$3 or $\sim$4.5\, kpc.

Based on the presented analysis, a low-magnetized neutron star located
in the Scutum--Crux spiral arm and covered by a uniformly emitting
carbon atmosphere seems to be the most plausible hypothesis for the \cco.

Recently, \citet{Fukuda:etal:14} argued that the remnant
\hessj should be associated with the 
3\,kpc expanding arm
based on their morphological studies of the $^{12}$CO and HI    
observations. 
The authors argued that 
the distance to the arm in the direction of the remnant is 
in the range 5.2--6.1\,kpc. 
Such a distance is formally still 
compatible with the carbon atmosphere models (Fig.\,\ref{fig:RMd}) but
requires the mass and radius of the compact star even higher than for
4.5\,kpc. On the other hand, in case of our hydrogen atmosphere
models, the suggested 
5.2--6.1\,kpc would lead to unrealistically low NS masses and radii
(Fig.\,\ref{fig:RMHatm}). The distance suggested by
\citet{Fukuda:etal:14} is thus not favored by our studies. This
possibility, however, cannot be rejected at this stage.

According to the current cooling theories, the \cco
can be considered as an isolated cooling NS which must have
been cooling down very slowly. For a reasonable assumption on the
source age of $\sim$10--40\,kyr, it is the hottest cooling NS
observed so far. It must have a very low neutrino luminosity   
{\em and} an unusually heat-transparent blanketing envelope. The required low 
neutrino luminosity can be realized in a star with strong proton superfluidity
in the core (where neutrino emission is produced by neutron-neutron
collisions). High heat transparency can be provided by the presence of
sufficient amount of carbon in the heat blanketing envelope. For a star's
age of 27\,kyr, the heat blanket should contain the maximum amount of carbon,
$\Delta M \sim 10^{-8}\,{\rm M_\odot}$. For an age of $\sim$10\,kyr,
proton superfluidity can be weaker and/or $\Delta M$ can be
smaller. In any case, the star should have
a rather large amount of carbon in the surface layers.
We demonstrated that the cooling theory has a potential
to put further constraints on the mass and radius of \xmmu. 
It also seems to disfavor larger distances to
the source, above $\sim$4--5\,kpc, as they lead to a larger age of the NS.

Generally, our analysis demonstrates that CCOs are very promising objects
for the measurements of NS masses and radii and, thus, for probing
the EOS of superdense matter. 
The increase in the total exposure time led to a significant
reduction of the allowed region in the mass-radius plane. 
Some constraints on mass and radius can also be given with only a
lower limit on the distance (Fig.\,\ref{fig:RMd}).
The obtained best-fit NS radius of 12.4\,km (for the preferred
distance of 3.2\,kpc) turns out to be
very close to the preferable NS radii of $11-13$\,km obtained 
with the current theoretical modeling of EOSs, the
experimental data from heavy ion collisions, and the observations
of some X-ray bursting NSs and the NSs in quiescent states of soft
X-ray transients \citep{Steiner:etal:13,Lattimer:Steiner:14}.  

As the future improvement of the observational data will further
reduce the statistical errors on mass and radius, the
uncertainties in the theoretical modeling
might become the dominant source of errors.
In this respect, we note the following physical processes not yet 
taken into account in our carbon NS atmosphere models which might
affect the shape of the emergent spectra: (i) deviations of carbon
ion number densities from LTE values 
\citep[see, e. g.,][]{Rauch:etal:08}, (ii) non-coherent electron
scattering (Compton effect), and (iii)
influence of the magnetic field $B\sim 10^{10} - 10^{11}$\,G. 

\begin{acknowledgements}
VS acknowledges the support by the German Research Foundation
(DFG) grant SFB/Transregio 7 "Gravitational Wave Astronomy" and
Russian Foundation for Basic Research (grant
12-02-97006-r-povolzhe-a). DY acknowledges
partial support by the Russian Foundation for Basic 
Research (grants No.~14-02-00868-a and 13-02-12017-ofi-M)
 and by the State Program ``Leading Scientific Schools of RF'' (grant
 NSh 294.2014.2).
\end{acknowledgements}

\bibliographystyle{aa}
\bibliography{refs}

\begin{thebibliography}{45}
\expandafter\ifx\csname natexlab\endcsname\relax\def\natexlab#1{#1}\fi

\bibitem[{{Abramowski} {et~al.}(2011){Abramowski}, {Acero}, {Aharonian},
  {Akhperjanian}, {Anton}, {Balzer}, {Barnacka}, {Barres de Almeida},
  {Becherini}, {Becker}, {Behera}, {Bernl{\"o}hr}, {Bochow}, {Boisson},
  {Bolmont}, {Bordas}, {Brucker}, {Brun}, {Brun}, {Bulik}, {B{\"u}sching},
  {Carrigan}, {Casanova}, {Cerruti}, {Chadwick}, {Charbonnier}, {Chaves},
  {Cheesebrough}, {Chounet}, {Clapson}, {Coignet}, {Cologna}, {Conrad},
  {Dalton}, {Daniel}, {Davids}, {Degrange}, {Deil}, {Dickinson},
  {Djannati-Ata{\"i}}, {Domainko}, {Drury}, {Dubois}, {Dubus}, {Dutson},
  {Dyks}, {Dyrda}, {Egberts}, {Eger}, {Espigat}, {Fallon}, {Farnier}, {Fegan},
  {Feinstein}, {Fernandes}, {Fiasson}, {Fontaine}, {F{\"o}rster},
  {F{\"u}{\ss}ling}, {Gallant}, {Gast}, {G{\'e}rard}, {Gerbig}, {Giebels},
  {Glicenstein}, {Gl{\"u}ck}, {Goret}, {G{\"o}ring}, {H{\"a}ffner}, {Hague},
  {Hampf}, {Hauser}, {Heinz}, {Heinzelmann}, {Henri}, {Hermann}, {Hinton},
  {Hoffmann}, {Hofmann}, {Hofverberg}, {Holler}, {Horns}, {Jacholkowska}, {de
  Jager}, {Jahn}, {Jamrozy}, {Jung}, {Kastendieck}, {Katarzy{\'n}ski}, {Katz},
  {Kaufmann}, {Keogh}, {Khangulyan}, {Kh{\'e}lifi}, {Klochkov}, {Klu{\'z}niak},
  {Kneiske}, {Komin}, {Kosack}, {Kossakowski}, {Laffon}, {Lamanna}, {Lennarz},
  {Lohse}, {Lopatin}, {Lu}, {Marandon}, {Marcowith}, {Masbou}, {Maurin},
  {Maxted}, {McComb}, {Medina}, {M{\'e}hault}, {Moderski}, {Moulin}, {Naumann},
  {Naumann-Godo}, {de Naurois}, {Nedbal}, {Nekrassov}, {Nguyen}, {Nicholas},
  {Niemiec}, {Nolan}, {Ohm}, {de O{\~n}a Wilhelmi}, {Opitz}, {Ostrowski},
  {Oya}, {Panter}, {Paz Arribas}, {Pedaletti}, {Pelletier}, {Petrucci}, {Pita},
  {P{\"u}hlhofer}, {Punch}, {Quirrenbach}, {Raue}, {Rayner}, {Reimer},
  {Reimer}, {Renaud}, {de los Reyes}, {Rieger}, {Ripken}, {Rob}, {Rosier-Lees},
  {Rowell}, {Rudak}, {Rulten}, {Ruppel}, {Ryde}, {Sahakian}, {Santangelo},
  {Schlickeiser}, {Sch{\"o}ck}, {Schulz}, {Schwanke}, {Schwarzburg},
  {Schwemmer}, {Sikora}, {Skilton}, {Sol}, {Spengler}, {Stawarz}, {Steenkamp},
  {Stegmann}, {Stinzing}, {Stycz}, {Sushch}, {Szostek}, {Tavernet}, {Terrier},
  {Tluczykont}, {Valerius}, {van Eldik}, {Vasileiadis}, {Venter}, {Vialle},
  {Viana}, {Vincent}, {V{\"o}lk}, {Volpe}, {Vorobiov}, {Vorster}, {Wagner},
  {Ward}, {White}, {Wierzcholska}, {Zacharias}, {Zajczyk}, {Zdziarski}, {Zech},
  \& {Zechlin}}]{HESS:2011}
{Abramowski}, A., {Acero}, F., {Aharonian}, F., {et~al.} 2011, \aap, 531, A81

\bibitem[{{Acero} {et~al.}(2009){Acero}, {P{\"u}hlhofer}, {Klochkov}, {Komin},
  {Gallant}, {Horns}, {Santangelo}, \& {for the
  H.~E.~S.~S.~Collaboration}}]{Acero:etal:09}
{Acero}, F., {P{\"u}hlhofer}, G., {Klochkov}, D., {et~al.} 2009, in Proc. of
  the 31th ICRC 2009, Lodz (arXiv e-Print 0907.0642)

\bibitem[{{Bamba} {et~al.}(2012){Bamba}, {P{\"u}hlhofer}, {Acero}, {Klochkov},
  {Tian}, {Yamazaki}, {Li}, {Horns}, {Kosack}, \& {Komin}}]{Bamba:etal:12}
{Bamba}, A., {P{\"u}hlhofer}, G., {Acero}, F., {et~al.} 2012, \apj, 756, 149

\bibitem[{{Bogdanov}(2014)}]{Bogdanov:14}
{Bogdanov}, S. 2014, ArXiv e-prints

\bibitem[{{de Jager} {et~al.}(1989){de Jager}, {Raubenheimer}, \&
  {Swanepoel}}]{Jager:etal:89}
{de Jager}, O.~C., {Raubenheimer}, B.~C., \& {Swanepoel}, J.~W.~H. 1989, in
  Data Analysis in Astronomy, ed. V.~{di Gesu}, L.~{Scarsi}, P.~{Crane}, J.~H.
  {Friedman}, S.~{Levialdi}, \& M.~C. {Maccarone}, 21

\bibitem[{{Fukuda} {et~al.}(2014){Fukuda}, {Yoshiike}, {Sano}, {Torii},
  {Yamamoto}, {Acero}, \& {Fukui}}]{Fukuda:etal:14}
{Fukuda}, T., {Yoshiike}, S., {Sano}, H., {et~al.} 2014, \apj, 788, 94

\bibitem[{{Gnedin} {et~al.}(2001){Gnedin}, {Yakovlev}, \&
  {Potekhin}}]{Gnedin:etal:01}
{Gnedin}, O.~Y., {Yakovlev}, D.~G., \& {Potekhin}, A.~Y. 2001, \mnras, 324, 725

\bibitem[{{Gotthelf} {et~al.}(2013){Gotthelf}, {Halpern}, \&
  {Alford}}]{Gotthelf:etal:13}
{Gotthelf}, E.~V., {Halpern}, J.~P., \& {Alford}, J. 2013, \apj, 765, 58

\bibitem[{{Haensel} {et~al.}(2007){Haensel}, {Potekhin}, \&
  {Yakovlev}}]{Haensel:etal:07}
{Haensel}, P., {Potekhin}, A.~Y., \& {Yakovlev}, D.~G., eds. 2007, Astrophysics
  and Space Science Library, Vol. 326, {Neutron Stars 1 : Equation of State and
  Structure}

\bibitem[{{Halpern} \& {Gotthelf}(2010{\natexlab{a}})}]{Halpern:Gotthelf:10c}
{Halpern}, J.~P. \& {Gotthelf}, E.~V. 2010{\natexlab{a}}, \apj, 725, 1384

\bibitem[{{Halpern} \& {Gotthelf}(2010{\natexlab{b}})}]{Halpern:Gotthelf:10}
{Halpern}, J.~P. \& {Gotthelf}, E.~V. 2010{\natexlab{b}}, \apj, 709, 436

\bibitem[{{Halpern} \& {Gotthelf}(2010{\natexlab{c}})}]{Halpern:Gotthelf:10:b}
{Halpern}, J.~P. \& {Gotthelf}, E.~V. 2010{\natexlab{c}}, \apj, 710, 941

\bibitem[{{Heinke} \& {Ho}(2010)}]{Heinke:Ho:10}
{Heinke}, C.~O. \& {Ho}, W.~C.~G. 2010, \apjl, 719, L167

\bibitem[{{Ho} \& {Heinke}(2009)}]{Ho:Heinke:09}
{Ho}, W.~C.~G. \& {Heinke}, C.~O. 2009, \nat, 462, 71

\bibitem[{{Hou} {et~al.}(2009){Hou}, {Han}, \& {Shi}}]{Hou:etal:09}
{Hou}, L.~G., {Han}, J.~L., \& {Shi}, W.~B. 2009, \aap, 499, 473

\bibitem[{{Hubeny} {et~al.}(1994){Hubeny}, {Hummer}, \&
  {Lanz}}]{Hubeny:etal:94}
{Hubeny}, I., {Hummer}, D.~G., \& {Lanz}, T. 1994, \aap, 282, 151

\bibitem[{{Hummer} \& {Mihalas}(1988)}]{Hummer:Mihalas:88}
{Hummer}, D.~G. \& {Mihalas}, D. 1988, \apj, 331, 794

\bibitem[{{Jansen} {et~al.}(2001){Jansen}, {Lumb}, {Altieri}, {Clavel}, {Ehle},
  {Erd}, {Gabriel}, {Guainazzi}, {Gondoin}, {Much}, {Munoz}, {Santos},
  {Schartel}, {Texier}, \& {Vacanti}}]{Jansen:etal:01}
{Jansen}, F., {Lumb}, D., {Altieri}, B., {et~al.} 2001, \aap, 365, L1

\bibitem[{{Karpova} {et~al.}(2014){Karpova}, {Danilenko}, {Shibanov},
  {Shternin}, \& {Zyuzin}}]{Karpova:etal:14}
{Karpova}, A., {Danilenko}, A., {Shibanov}, Y., {Shternin}, P., \& {Zyuzin}, D.
  2014, \apj, 789, 97

\bibitem[{{Kirichenko} {et~al.}(2014){Kirichenko}, {Danilenko}, {Shibanov},
  {Shternin}, {Zharikov}, \& {Zyuzin}}]{Kirichenko:etal:14}
{Kirichenko}, A., {Danilenko}, A., {Shibanov}, Y., {et~al.} 2014, \aap, 564,
  A81

\bibitem[{{Klochkov} {et~al.}(2013){Klochkov}, {P{\"u}hlhofer}, {Suleimanov},
  {Simon}, {Werner}, \& {Santangelo}}]{Klochkov:etal:13}
{Klochkov}, D., {P{\"u}hlhofer}, G., {Suleimanov}, V., {et~al.} 2013, \aap,
  556, A41

\bibitem[{{Lattimer} \& {Prakash}(2007)}]{Lattimer:Prakash:07}
{Lattimer}, J.~M. \& {Prakash}, M. 2007, \physrep, 442, 109

\bibitem[{{Lattimer} \& {Steiner}(2014)}]{Lattimer:Steiner:14}
{Lattimer}, J.~M. \& {Steiner}, A.~W. 2014, \apj, 784, 123

\bibitem[{{Marelli} {et~al.}(2013){Marelli}, {De Luca}, {Salvetti}, {Sartore},
  {Sartori}, {Caraveo}, {Pizzolato}, {Saz Parkinson}, \&
  {Belfiore}}]{Marelli:etal:13}
{Marelli}, M., {De Luca}, A., {Salvetti}, D., {et~al.} 2013, \apj, 765, 36

\bibitem[{{Page}(1993)}]{Page:93}
{Page}, D. 1993, in Nuclear Physics in the Universe, ed. M.~W. {Guidry} \&
  M.~R. {Strayer}, 151--162

\bibitem[{{Page} {et~al.}(2006){Page}, {Geppert}, \& {Weber}}]{Page:etal:06}
{Page}, D., {Geppert}, U., \& {Weber}, F. 2006, Nuclear Physics A, 777, 497

\bibitem[{{Page} {et~al.}(2009){Page}, {Lattimer}, {Prakash}, \&
  {Steiner}}]{Page:etal:09}
{Page}, D., {Lattimer}, J.~M., {Prakash}, M., \& {Steiner}, A.~W. 2009, \apj,
  707, 1131

\bibitem[{{Pavlov} \& {Luna}(2009)}]{Pavlov:Luna:09}
{Pavlov}, G.~G. \& {Luna}, G.~J.~M. 2009, \apj, 703, 910

\bibitem[{{Pavlov} {et~al.}(2002){Pavlov}, {Sanwal}, {Garmire}, \&
  {Zavlin}}]{Pavlov:etal:02}
{Pavlov}, G.~G., {Sanwal}, D., {Garmire}, G.~P., \& {Zavlin}, V.~E. 2002, in
  Astronomical Society of the Pacific Conference Series, Vol. 271, Neutron
  Stars in Supernova Remnants, ed. P.~O. {Slane} \& B.~M. {Gaensler}, 247

\bibitem[{{Pavlov} {et~al.}(2004){Pavlov}, {Sanwal}, \&
  {Teter}}]{Pavlov:etal:04}
{Pavlov}, G.~G., {Sanwal}, D., \& {Teter}, M.~A. 2004, in IAU Symposium, Vol.
  218, Young Neutron Stars and Their Environments, ed. {F.~Camilo \&
  B.~M.~Gaensler}, 239

\bibitem[{{Posselt} {et~al.}(2013){Posselt}, {Pavlov}, {Suleimanov}, \&
  {Kargaltsev}}]{Posselt:etal:13}
{Posselt}, B., {Pavlov}, G.~G., {Suleimanov}, V., \& {Kargaltsev}, O. 2013,
  \apj, 779, 186

\bibitem[{{Protheroe}(1987)}]{Protheroe:etal:87}
{Protheroe}, R.~J. 1987, Proceedings of the Astronomical Society of Australia,
  7, 167

\bibitem[{{Rauch} {et~al.}(2008){Rauch}, {Suleimanov}, \&
  {Werner}}]{Rauch:etal:08}
{Rauch}, T., {Suleimanov}, V., \& {Werner}, K. 2008, \aap, 490, 1127

\bibitem[{{Steiner} {et~al.}(2013){Steiner}, {Lattimer}, \&
  {Brown}}]{Steiner:etal:13}
{Steiner}, A.~W., {Lattimer}, J.~M., \& {Brown}, E.~F. 2013, \apjl, 765, L5

\bibitem[{{Str{\"u}der} {et~al.}(2001){Str{\"u}der}, {Briel}, {Dennerl},
  {Hartmann}, {Kendziorra}, {Meidinger}, {Pfeffermann}, {Reppin}, {Aschenbach},
  {Bornemann}, {Br{\"a}uninger}, {Burkert}, {Elender}, {Freyberg}, {Haberl},
  {Hartner}, {Heuschmann}, {Hippmann}, {Kastelic}, {Kemmer}, {Kettenring},
  {Kink}, {Krause}, {M{\"u}ller}, {Oppitz}, {Pietsch}, {Popp}, {Predehl},
  {Read}, {Stephan}, {St{\"o}tter}, {Tr{\"u}mper}, {Holl}, {Kemmer}, {Soltau},
  {St{\"o}tter}, {Weber}, {Weichert}, {von Zanthier}, {Carathanassis}, {Lutz},
  {Richter}, {Solc}, {B{\"o}ttcher}, {Kuster}, {Staubert}, {Abbey}, {Holland},
  {Turner}, {Balasini}, {Bignami}, {La Palombara}, {Villa}, {Buttler},
  {Gianini}, {Lain{\'e}}, {Lumb}, \& {Dhez}}]{Strueder:etal:01}
{Str{\"u}der}, L., {Briel}, U., {Dennerl}, K., {et~al.} 2001, \aap, 365, L18

\bibitem[{{Suleimanov} \& {Werner}(2007)}]{Suleimanov:Werner:07}
{Suleimanov}, V. \& {Werner}, K. 2007, \aap, 466, 661

\bibitem[{{Suleimanov} {et~al.}(2014){Suleimanov}, {Klochkov}, {Pavlov}, \&
  {Werner}}]{Suleimanov:etal:14}
{Suleimanov}, V.~F., {Klochkov}, D., {Pavlov}, G.~G., \& {Werner}, K. 2014,
  \apjs, 210, 13

\bibitem[{{Tian} {et~al.}(2008){Tian}, {Leahy}, {Haverkorn}, \&
  {Jiang}}]{Tian:etal:08}
{Tian}, W.~W., {Leahy}, D.~A., {Haverkorn}, M., \& {Jiang}, B. 2008, \apjl,
  679, L85

\bibitem[{{Tian} {et~al.}(2010){Tian}, {Li}, {Leahy}, {Yang}, {Yang},
  {Yamazaki}, \& {Lu}}]{Tian:etal:10}
{Tian}, W.~W., {Li}, Z., {Leahy}, D.~A., {et~al.} 2010, \apj, 712, 790

\bibitem[{{Turner} {et~al.}(2001){Turner}, {Abbey}, {Arnaud}, {Balasini},
  {Barbera}, {Belsole}, {Bennie}, {Bernard}, {Bignami}, {Boer}, {Briel},
  {Butler}, {Cara}, {Chabaud}, {Cole}, {Collura}, {Conte}, {Cros}, {Denby},
  {Dhez}, {Di Coco}, {Dowson}, {Ferrando}, {Ghizzardi}, {Gianotti}, {Goodall},
  {Gretton}, {Griffiths}, {Hainaut}, {Hochedez}, {Holland}, {Jourdain},
  {Kendziorra}, {Lagostina}, {Laine}, {La Palombara}, {Lortholary}, {Lumb},
  {Marty}, {Molendi}, {Pigot}, {Poindron}, {Pounds}, {Reeves}, {Reppin},
  {Rothenflug}, {Salvetat}, {Sauvageot}, {Schmitt}, {Sembay}, {Short},
  {Spragg}, {Stephen}, {Str{\"u}der}, {Tiengo}, {Trifoglio}, {Tr{\"u}mper},
  {Vercellone}, {Vigroux}, {Villa}, {Ward}, {Whitehead}, \&
  {Zonca}}]{Turner:etal:01}
{Turner}, M.~J.~L., {Abbey}, A., {Arnaud}, M., {et~al.} 2001, \aap, 365, L27

\bibitem[{{Weisskopf} {et~al.}(2011){Weisskopf}, {Tennant}, {Yakovlev},
  {Harding}, {Zavlin}, {O'Dell}, {Elsner}, \& {Becker}}]{Weisskopf:etal:11}
{Weisskopf}, M.~C., {Tennant}, A.~F., {Yakovlev}, D.~G., {et~al.} 2011, \apj,
  743, 139

\bibitem[{{Yakovlev} {et~al.}(2011){Yakovlev}, {Ho}, {Shternin}, {Heinke}, \&
  {Potekhin}}]{Yakovlev:etal:11}
{Yakovlev}, D.~G., {Ho}, W.~C.~G., {Shternin}, P.~S., {Heinke}, C.~O., \&
  {Potekhin}, A.~Y. 2011, \mnras, 411, 1977

\bibitem[{{Yakovlev} \& {Pethick}(2004)}]{Yakovlev:Pethick:04}
{Yakovlev}, D.~G. \& {Pethick}, C.~J. 2004, \araa, 42, 169

\bibitem[{{Zavlin} {et~al.}(1998){Zavlin}, {Pavlov}, \&
  {Trumper}}]{Zavlin:etal:98}
{Zavlin}, V.~E., {Pavlov}, G.~G., \& {Trumper}, J. 1998, \aap, 331, 821

\bibitem[{{Zavlin} {et~al.}(1999){Zavlin}, {Tr{\"u}mper}, \&
  {Pavlov}}]{Zavlin:etal:99}
{Zavlin}, V.~E., {Tr{\"u}mper}, J., \& {Pavlov}, G.~G. 1999, \apj, 525, 959

\end{thebibliography}

\end{document}